\documentclass[prx,twocolumn,amsmath,amssymb,notitlepage]{revtex4-1}
\usepackage{tikz}
\usepackage{circuitikz}
\usetikzlibrary{arrows,shapes.gates.logic.US,shapes.gates.logic.IEC,calc}
\usepackage{graphicx}
\usepackage{amssymb}
\usepackage{amsfonts}
\usepackage{amsmath}
\usepackage{amsbsy}
\usepackage{natbib}
\usepackage{comment}
\usepackage{amssymb}
\usepackage{color}
\usepackage{float}
\DeclareMathAlphabet\mathbfcal{OMS}{cmsy}{b}{n}
\DeclareMathAlphabet{\mathpzc}{OT1}{pzc}{m}{it}
\usepackage[colorlinks=true, urlcolor=blue, citecolor=blue, linkcolor=blue]{hyperref}
\usepackage{bm}
\usepackage{bm}
\usepackage{bbold}

\renewcommand{\vec}[1]{\mbox{\boldmath$\mathrm{#1}$}}
\let\sb=_ \catcode`\_=\active \def_#1{\ensuremath \sb{\rm#1}}

\renewcommand{\vec}[1]{\mbox{\boldmath$\mathrm{#1}$}}
\newcommand{\be}{\begin{equation}}
\newcommand{\ee}{\end{equation}}
\newcommand{\ben}{\begin{eqnarray}}
\newcommand{\een}{\end{eqnarray}}

\newcommand{\beginsupplement}{%
	\setcounter{equation}{0}
	\renewcommand{\theequation}{S\arabic{equation}}%
	\setcounter{figure}{0}
	\renewcommand{\thefigure}{S\arabic{figure}}%
}

\begin{document}


\title{Skyrmion Echo in a system of interacting Skyrmions}

\author{X.-G. Wang$^{1}$,  Guang-hua Guo$^{1}$, A. Dyrda\l$^{2}$,
  J. Barna\'s$^{2}$, V. K. Dugaev$^3$, S. S. P. Parkin$^{4}$, A. Ernst$^{4,5}$ and L. Chotorlishvili$^3$}
\address{$^1$ School of Physics and Electronics, Central South University, Changsha 410083, China \\
$^2$ Faculty of Physics, Adam Mickiewicz University, ul. Uniwersytetu Pozna\'nskiego 2, 61-614 Pozna\'n, Poland\\
$^3$ Department of Physics and Medical Engineering, Rzesz\'ow University of Technology, 35-959 Rzesz\'ow, Poland\\
$^4$ Max Planck Institute of Microstructure Physics, Weinberg 2, D-06120 Halle, Germany\\
$^{5}$ Institute for Theoretical Physics, Johannes Kepler University, Altenberger Stra\ss e 69, 4040 Linz, Austria
}
\date{\today}
\begin{abstract} We consider helical rotation of skyrmions confined in
  the potentials formed by nano-disks. Based on numerical and
  analytical calculations we propose the skyrmion echo phenomenon. The physical mechanism of the skyrmion echo
  formation is also proposed. Due to the distortion of the lattice,
  impurities, or pinning effect, confined skyrmions experience
  slightly different local fields, which leads to dephasing of the
  initial signal.  The interaction between skyrmions also can
  contribute to the dephasing process. However, switching the
  magnetization direction in the nanodiscs (e.g. by spin transfer
  torque) also switches the helical rotation of the skyrmions from
  clockwise to anticlockwise (or vice-versa), and this restores the
  initial signal (which is the essence of skyrmion
  echo).  \end{abstract}

\maketitle

\begin{figure}
	\includegraphics[width=0.45\textwidth]{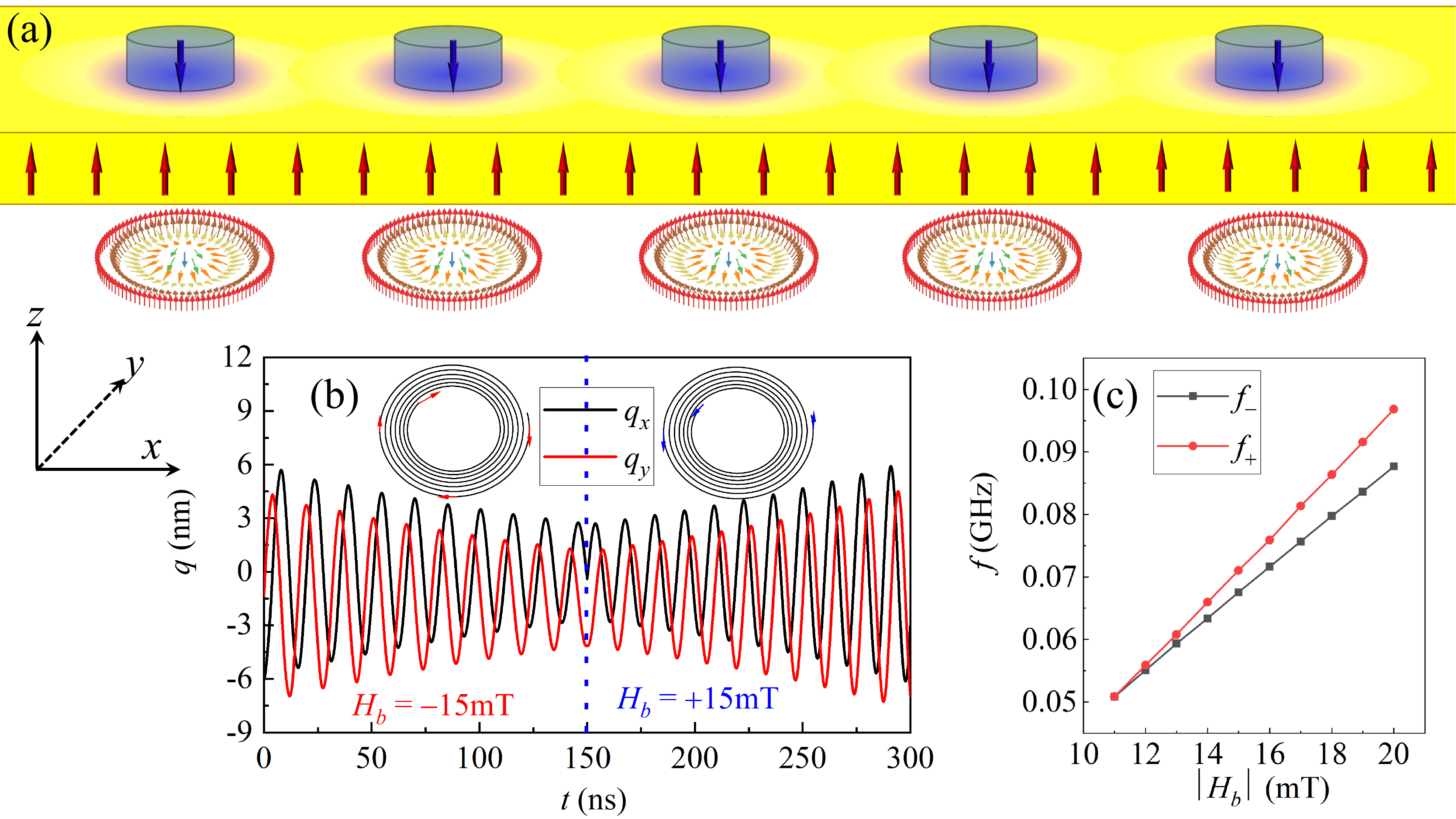}
	\caption{\label{emodel} (a) Schematic of a chain of pinned skyrmions.
          Skyrmions are created in a thin magnetic film, on
          top of which there are magnetic nanodisks. The blue arrows
          show the direction of the field exerted on the skyrmions by
          the nanodisks.  The skyrmions are confined in the regions
          below the nanodisks. Initially a particular skyrmion is
          embedded in the vicinity of the center of the corresponding
          nanodisk. (b) The skyrmion dynamics is described in terms
          of the position of the skyrmion center $ (q_x, q_y)$ with
          respect to the center of the corresponding nanodisk. For a
          negative field $ H_b =-15 $ mT, the skyrmion moves along the
          helix trajectory towards the center of the nanodisk
          $ (0, 0) $. Upon the sudden quenching of the field $ H_b $ from
          $ -15 $ mT to $ 15 $ mT, at $ t = 150 $ ns, the
          skyrmion moves along the helix trajectory away from the
          center of the nanodisk. (c) Frequencies of the clockwise
          ($ f_{-} $ , $ H_b < 0 $) and anticlockwise ($ f_{+}
          $,$
          H_b > 0 $)
          precessions as a function of the amplitude of the field
          exerted on the skyrmion by the nanodisk, $ |H_b| $. The
          anticlockwise precession frequency is larger than that of
          the clockwise one.}
\end{figure}

\textit{Introduction} In 1950 Erwin Hahn discovered the effect that is
now known as the spin echo \cite{abragam1961principles}.  Due to the
inhomogeneity of a local magnetic field in solids, nuclear (or
electron) spins precess with slightly different
frequencies. Therefore, an initially excitated pulse decays after a
specific time. However, the application of a properly designed pulse
reverses the precession direction from clockwise to anticlockwise (or
vice-versa),
and this restores the initial signal. The dephasing/rephasing
mechanism of precessing spins was subsequently explained by Bloom
\cite{PhysRev.98.1105}. In this letter, we show that the echo mechanism is
  applicable not only to precessing spins but also to more complex
  objects, such as, for instance, skyrmions. Skyrmions are topological
  solitons discovered in non-Abelian gauge field theories
  \cite{schroers1995bogomol,samoilenka2017gauged,battye2013isospinning,jennings2014broken},
  and subsequently in condensed matter physics
  \cite{seki2012observation,wilson2014chiral}. They have localized
  robust shapes and additionally possess a topological charge -- the
  conserved quantity underlying their topological protection. Magnetic
  films without inversion symmetry can host skyrmions and specific
  skyrmionic magnetic textures described by the local magnetization
  $\textbf{M}(\textbf{r})$ have been studied theoretically and then
  have been discovered experimentally. Owing to their topological
  properties and potential applications, skyrmions are currently of
  great interest, both theoretical and experimental. Skyrmions can exists as independent objects, but can
  also form regular skyrmion lattices (i.e. skyrmion crystals). For
  fundamental aspects of skyrmions, we refer to classical handbooks
  \cite{rajaraman1982solitons,altland2010condensed}.  Concerning
  modern mathematical aspects of skyrmions we refer to
  \cite{barton2020magnetic}.  The key source of the skyrmion formation and non-colinear magnetic textures is either 
  the interfacial Dzyaloshinskii–Moriya interaction (DMI)
  \cite{white2014electric,derras2018quantum, haldar2018first,
psaroudaki2017quantum,van2013magnetic,rohart2016path,tsesses2018optical,wang2018electric,wang2019thermally},
 competition between ferromagnetic and antiferromagnetic exchange interactions \cite{leonov2015multiply}, or bulk
DMI in case of antiferromagnetic skyrmions \cite{Zhang2016,PhysRevLett.116.147203}.
Individual skyrmions can be pinned by specific confinement potentials, e.g., those created by inhomogeneous magnetic/electric field \cite{Wang2017,wang2020optical}, defects \cite{Hanneken2016}, spin transfer torque \cite{Ma2015}, or magnetic
  nanodisks \cite{PhysRevB.95.174416,PhysRevLett.110.167201}.
  Recent interest in skyrmionics is focuses especially on potential
  applications of skyrmions in data storage and processing technologies
  \cite{Tomasello2014, doi:10.1063/5.0042917,Zzvorka2019,Zhang2015,Fert2013,Wiesendanger2016}.
   In this letter, we explore the formation of the skyrmion echo in a system of interacting skyrmions.
In Fig. \ref{emodel}(a) we present the model considered in this
paper. The bottom magnetic thin film hosts several skyrmions. On top
of this film there are magnetic nanodisks that confine skyrmions in
the regions below these nanodisks. Being confined the skyrmions may
perform circular clockwise or anticlockwise motion in these regions,
and the winding direction depends on the confined field from the
nanodisks. By reversing this field one can switch the direction of the
skyrmion winding trajectories, and this behavior can play a role
similar to that of the second (rephasing) pulse in the Hahn's spin
echo case. Thus, due to dephasing induced by inhomogeneities in the
confining magnetic field, one may observe the skyrmion echo -- a
phenomenon similar to the spin echo. The inhomogeneous
field experienced by skyrmions may be formed in various ways, e.g.,
from lattice distortions or pinning sites. Skyrmion-skyrmion
interactions can also contribute to the dephasing process. Material parameters influence the shape and width of the
resonance spectra. Consequently, from the skyrmion echo one can
extract information on the host material and on the skyrmion-skyrmion
interactions. To describe the process, we exploit the Landau-Lifshitz-Gilbert (LLG)
equation:
\begin{equation}
\displaystyle \frac{\partial \vec{M}}{\partial t} = - \gamma \vec{M} \times \vec{H}_{\mathrm{\rm eff}} + \frac{\alpha}{M_{s}} \vec{M} \times \frac{\partial \vec{M}}{\partial t}.
\label{LLG}
\end{equation}
Here, $ \vec{M} = M_s \vec{m} $, with $ M_s$ and $\vec{m}$ being the saturation magnetization and unit vector along the
magnetization $\vec{M}$, while $ \alpha $ is the phenomenological Gilbert damping
constant.  The total effective magnetic field,
$ \vec{H}_{eff}$,  reads:
$\vec{H}_{eff}= \frac{2 A_{ex}}{\mu_0 M_{s}} \nabla^2 \vec{m} + H_z
\vec{z} + H_b \vec{z} -\frac{1}{\mu_0 M_s} \frac{\delta E_D}{\delta
  \vec{m}} $,
where the first term describes the internal exchange field with the
exchange stiffness $ A_{ex} $, the second term corresponds to the
external magnetic field $ H_z $ ($\vec{z}$ is a unit vector along the
axis $z$ normal to the film), the third term specifies coupling
between the thin film and nanodisks, while the last term is the DM
field, with the DM interaction energy density
$ E_D = D_m [(m_z \frac{d m_x}{dx} - m_x \frac{dm_z}{dx}) + (m_z
\frac{d m_y}{dy} - m_y \frac{dm_z}{dy})] $
and $ D_m $ being the strength of the DM interaction. The field $H_b$ is due to all nanodisks and is nonuniform
as contributions from different nanodisks may be different in
magnitude. We assume that an individual (say $i$-th) nanodisk creates
the confined field acting only on a single skyrmion, $H_b=H_b^i$. The
index $i$ will be omitted in general, and will be included only where
necessary.  In the supplementary information \cite{supp} we show that the main
  conclusions concerning the skyrmion echo are still valid when the
  anisotropy and demagnetization fields are included into consideration.
In numerical calculations we adopt typical parameters
that correspond to Co/heavy-metal multi-layers:
$ A_{ex} = 10 \rm{pJ}/\rm{m} $, $ D_m = 0.2 \rm{mJ}/\rm{m}^{2} $, and
$ M_s = 1.2\times 10^6 \rm{A}/\rm{m}$.  The bias magnetic field
$ H_z = 100 $ mT is used for stabilization of the skyrmion
structure. The skyrmion width is 45 nm (see
  supplementary information \cite{supp}). The size of the ferromagnetic layer is
$ 3000 \times 240 \times 3 \rm{nm}^3$. The ferromagnetic layer is discretized with the cells of size
$ 3 \times 3 \times 3 \rm{nm}^3 $. The radius of the upper nanodisks
is $ 12 $ nm. These nanodisks exert a coupling field $ H_b \vec{z} $
on the skyrmions in the bottom thin ferromagnetic film.
  The distance between neighboring nanodiskes is $ 270 $ nm.

\textit{Results}: Let us consider now the dynamics of a skyrmion
located in the region below a specific nanodisk. This dynamics is
described by the position of the skyrmion center $ (q_x, q_y)$ with
respect to the center of the corresponding nanodisk, at $(0,0)$.
 Assume the skyrmion was initially slightly displaced
  from the center of the nanodisk. When the field exerted by the
nanodisk on the skyrmion is negative, $ H_b < 0 $, the skyrmion moves
then along the helix trajectory winding clockwise about the nanodisk
center, see Fig. \ref{emodel}(b). The corresponding precession
frequency is approximately equal to 0.068 GHz. Upon reversing
direction of the field $ H_b $ to $ H_b > 0 $, the skyrmion starts to
move gradually away from the nanodisk center with the anticlockwise
precession. Interestingly, the frequency of an anticlockwise
precession (0.071 GHz) is slightly larger than that of the clockwise
one (0.068 GHz). This asymmetry in precession under the opposite
magnetic fields $ H_b $ and $-H_b $ is confirmed by detailed
calculations for different amplitudes of $ H_b $, see
Fig. \ref{emodel}(c).

An interesting question is how fast the system responds to the
switching between clockwise and anticlockwise regimes. To explore this
problem,
we 
quench the sign of the exerted field $ H_b $ from $ -15 $ mT to $ 15 $
mT.  Such a quench can be achieved for instance by reversing the
magnetization orientation in the nanodisk, e.g., via a strong
spin-transfer torque. We find that the skyrmion reacts almost
instantaneously as follows from Fig. \ref{emodel}(b). \vspace{0.1cm}\\
We elaborate the experimentally feasible scheme for a
skyrmion echo. For the spin echo, the spins
initially aligned parallel to the $z$ axis are rotated by a $\pi/2$
pulse applied at $t=0$, and then they start to precess in the $(xy)$
plane with different Larmor frequencies. The mismatch between these
frequencies leads to a dephasing of the precessions of different
spins. The second $\pi$ pulse applied at $t=\tau_0$ reverses the
precession direction and rephases the signal. In particular, the
refocusing of the spin orientations occurs at $t=2\tau_0$.
\begin{figure}
  \includegraphics[width=0.45\textwidth]{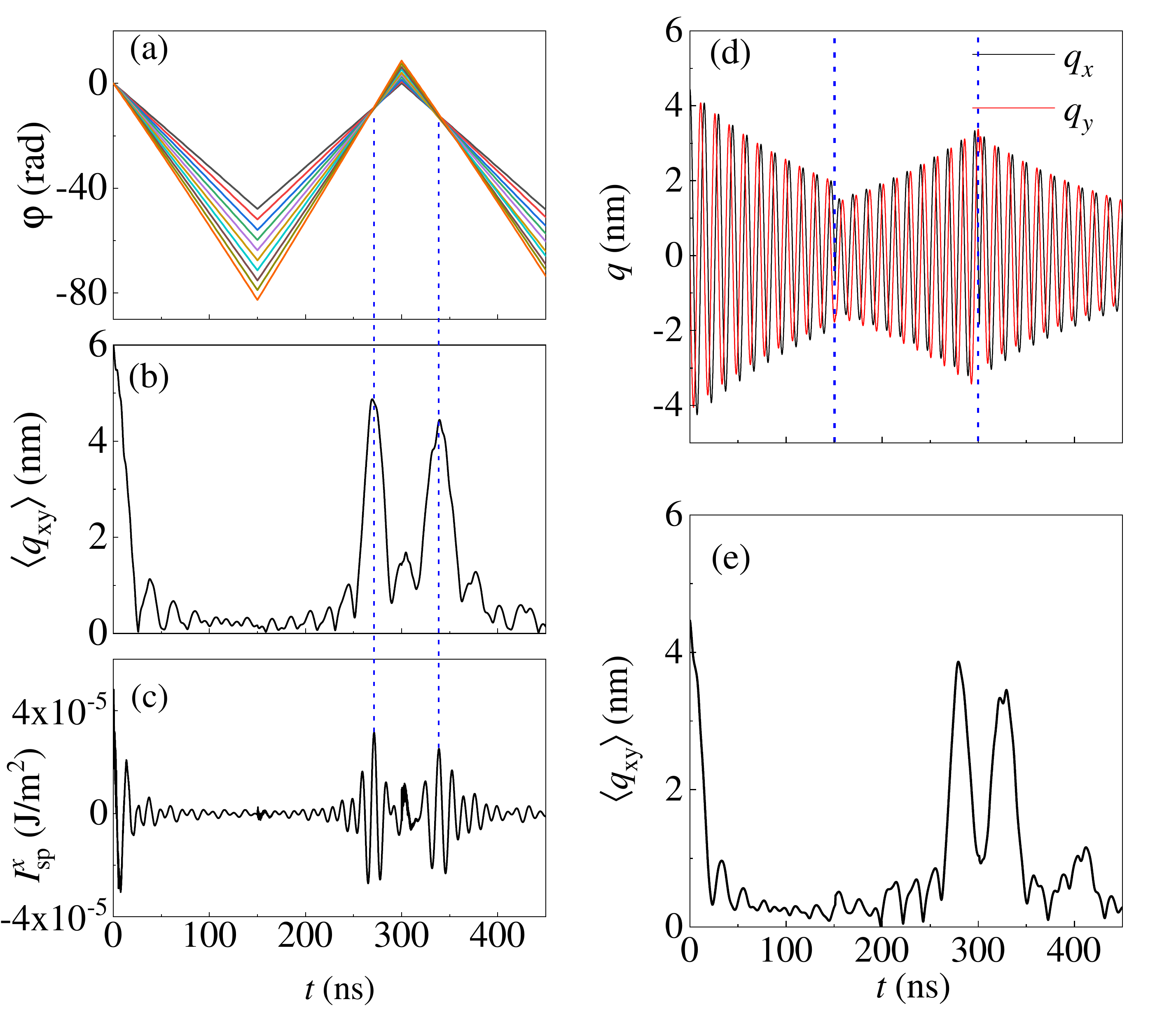}
  \caption{\label{echo} (a) Evolution of the precession phases
    $ \varphi_i = \arctan (q_y^i/q_x^i) $ for 10 skyrmions. After the
    time interval $ 150 $ ns, the direction of magnetic field $ H_b $
    in all nanodisks is reversed periodically. (b) The time dependence
    of the amplitude of skyrmion echo, $ \langle q_{xy} \rangle = \frac{1}{n} [(\sum_{i=1}^{n}q_x^i)^2 + (\sum_{i=1}^{n}q_y^i)^2 ]^{1/2} $ (here the average is over all $ n=10 $) skyrmions. Blue dashed lines mark the position of
    the skyrmion echo. (c) The time dependence of the $x$ component of
    the spin-pumping current $ I_{sp}^{x} $ into an additional heavy
    metal layer adjacent to the magnetic layer. (d-e) Skyrmion echo induced by current pulse. Dynamics of the
    upper nanodisks is included {\it via} the spin transfer torque
    due to a current pulse in an additional magnetic layer adjacent on
    top of the layer of nanodisks (saturation magnetization
    $ M_{s} = 1.4 \times 10^6 \rm{A}/\rm{m} $, exchange constant
    $ A_{ex} = 30 \rm{pJ}/\rm{m} $, and uniaxial anisotropy along
    $ z $ with constant $ K_{z} = 1 \times 10^6 $ J/m$ ^3 $). After a
    time interval $=150 $ ns, we periodically reverse the
    magnetization direction of the nanodisks {\it via} a strong spin
    transfer torque, and plot (d) the evolution of  $x$
      and $ y $ coordinates $q_x$ and
      $q_y$ of the skyrmion position (single skyrmion is plotted while the induced reversal in skyrmion precession applies to all 10 skyrmions), and (e) the averaged amplitude $ \langle q_{xy} \rangle $ of all 10 skyrmions.}
\end{figure}
Inspired by the idea of the spin echo, we generate a system of $n=10$
separated skyrmions, that are pinned under the ten nanodisks, see
Fig.~\ref{emodel}. The coupling fields from these nanodisks are
slightly different, inducing different precession frequencies of the
ten skyrmions. This can be achieved by a
  slight variation of the spacer thickness between the nanodisks and
  magnetic film, as the coupling strength is a function of the spacer
  thickness \cite{bruno1999theory}. In the model above, the pinning field induced by nanodisk is uniformly distributed and its amplitude varies from 10 mT to 20 mT.  Variation of the spacer thickness in the range of  1 nm is sufficient to generate such a change in the pinning field \cite{bruno1999theory, PhysRevLett.72.764}. All skyrmions are initially
steered away from the centers of nanodisks in the $ -x $
direction. This can be achieved by applying a spin-transfer torque or
a nonuniform magnetic field (for details see the
  supplementary information \cite{supp}). At the initial moment, $ t = 0 $, we
simultaneously release all the skyrmions. Different coupling fields
lead to dephasing of the skyrmion oscillations, as demonstrated in
Fig. \ref{echo}(a). At $ t = 150 $ ns, we abruptly reverse the
magnetization orientation in all the nanodisks. The skyrmion precessions
are then simultaneously reversed, and
re-phasing of the signal is achieved at $ t = 270 $ ns (this is the
skyrmion echo). The skyrmion echo signal can be seen clearly in
Fig. \ref{echo}(b), where the average position of the skyrmions along
the $x$-axis is shown.  It is evident that the skyrmion echo signal
occurs before $2 \tau_0 = 300$ ns. The reason for this is the
asymmetry between the clockwise and anticlockwise precession
frequencies, see Fig. \ref{emodel}(c).  After the decay of the
skyrmion echo signal, we reverse once again the field $H_b$ for all
the skyrmions at $ t = 300 $ns. The second signal of the skyrmion echo
now occurs after a shorter time, i.e. at $ t = 340 $ ns, see
Fig. \ref{echo}(a,b).  Apart from the
clockwise-anticlockwise precession asymmetry, the trajectory of the
skyrmion is not circular but spiral. Since the initial phases are set
to zero and the initial positions of the skyrmions are along the
$ -\textbf{x} $ direction $ (\lbrace-q_{\rm max}\rbrace, 0) $, the
first echo corresponds to the precession phases about $ - \pi $, and
all skyrmions situated along the $ +\textbf{x} $ direction
$ (\lbrace q_{\rm max}\rbrace, 0) $. In turn, the second echo is
characterized by the precession phases of $ -1.9 \pi $, and all
skyrmions set at
$(\lbrace-q_{\rm max}\rbrace\cos(0.1 \pi), \lbrace q_{\rm
  max}\rbrace\sin(0.1 \pi)) $.
The asymmetry in the dephasing and rephasing frequencies
  leads to a deviation from a perfect echo, i.e., the phase
  differences between 10 skyrmions are not precisely equal to
  zero. The skyrmion echo signal occurs at the minimum phase
  differences for the bias field $ H_z = 180 $ mT (see the
  supplementary information \cite{supp}).

The skyrmion echo can be detected experimentally by attaching a heavy
metal layer (for example a thin Pt layer) below the magnetic layer and
exploiting the spin pumping and inverse spin Hall effects. The
coherent precession of skyrmions pumps a spin current along the
$ \vec{z} $ axis towards the attached metallic layer,
$ \vec{I}_{\rm sp} = \frac{\hbar g_r}{ 4 \pi} (\vec{m} \times
\frac{\partial \vec{m}}{\partial t}) $.
The relevant spin mixing conductance is assumed to be
$ g_r = 7 \times 10^{18} \rm{m}^{-2} $. By means of the inverse spin
Hall effect, the spin pumping current is converted into an electric
current (voltage)
$ \vec_{J}_{\rm SH} = -\theta_{\rm SH} \frac{e}{\hbar}(\vec{I}_{\rm
  sp} \times \vec{z}) $,
where $ \theta_{SH} $ is the spin-Hall angle. The $ x $ component of
$ \vec{I}_{\rm sp} $ (that generates the $y$ component of
$ \vec_{J}_{\rm SH} $) in the bottom magnetic layer is shown in
Fig. \ref{echo}(c), where one can clearly see the skyrmion echo
signal.
\begin{figure}
  \includegraphics[width=0.45\textwidth]{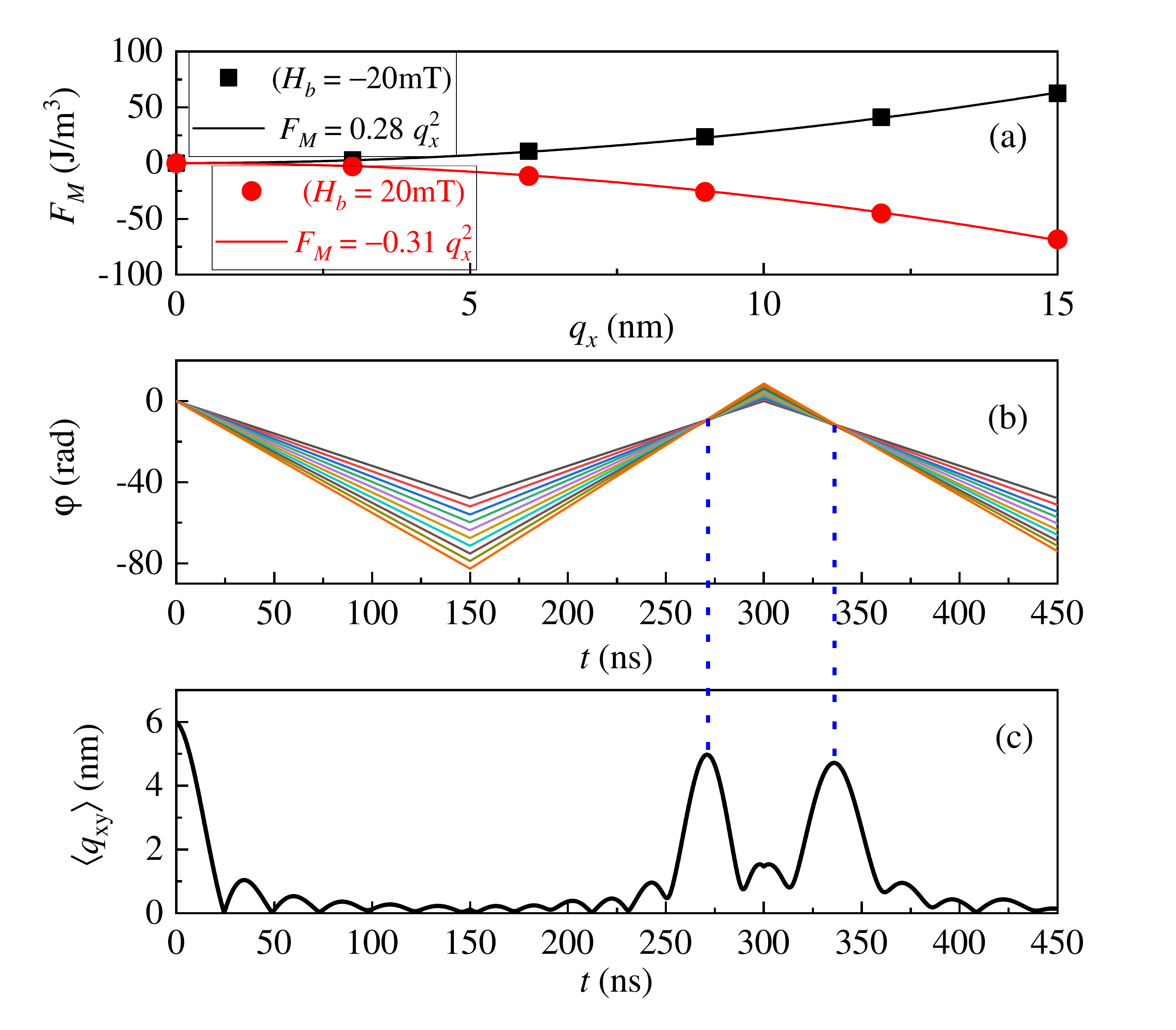}
  \caption{\label{thiele} (a) Averaged (over skyrmion area) magnetic free energy
    density $ F_M $ of the skyrmion as a function of the skyrmion position $ q_x $ with respect to the 
    corresponding nanodisk center, $ q_y = 0 $. The magnetic free energy is
    extracted from the simulation results, and the data  are
    fitted to the function  $ F_M = A r^2 $, where $ r^2 = q_x^2 $ + $ q_y^2 $. (b) and (c) Time evolution
    of the precession phases (b) and amplitude $ \langle q_{xy} \rangle = \frac{1}{n} [(\sum_{i=1}^{n}q_x^i)^2 + (\sum_{i=1}^{n}q_y^i)^2 ]^{1/2} $ of 10 skyrmions (c),
    calculated from the Thiele's equation.}
\end{figure}
To prove that the sign of the field $ H_b $ exerted on
the skyrmions can be reversed in experiment, we analyse the magnetic
dynamics of nanodisks due to a spin-transfer torque induced by a current
pulse in an additional layer above the nanodisks. This dynamics is
described by the LLG equation with the spin-transfer torque term
included. Coupling between the additional layer and a particular
nanodisk is introduced by the interlayer coupling field
$ H_{c} = \frac{J_c}{\mu_0 M_s t} \vec{m}$, and a different coupling
strength $ J_c $ is assumed for each of the nanodisks. Here $t$ and
$\vec{m}$ stand for the layer thickness and unit vector along the
nanodisk magnetization, respectively.
The corresponding numerical results are shown in Fig. \ref{echo}(d-e).
The magnetization of the nanodisks is reversed through the transfer torque
applied to the nanodisks from the adjacent layer,
$ \vec{\tau} = O_j \vec{m} \times \vec{p} \times \vec{m} $, where
$ \vec{p} = \vec{z} $ is the spin polarization orientation of the
electrons.  The spin transfer torque strength
$ O_j = \frac{\gamma P \hbar J_e}{ 2 \mu_0 e t M_s} $ is determined by
the electric current $ J_e $, thickness $ t = 2 $ nm, and current
polarization degree $ P = 0.5 $. We applied an ultrashort pulse of
amplitude $ J_e = 1.7 \times 10^7 \rm{A}/\rm{cm}^{2} $ and duration
of 2 ns.  Numerical results plotted in Fig. \ref{echo}(d-e) show the
skyrmion echo signal, and, thus, indirectly also confirm the switching of
the magnetization in the nanodisks.\vspace{0.1cm}\\
The skyrmion echo can also be described by Thiele's equations \cite{seidel,Iwasaki2013,Tomasello2014,PhysRevB.89.241101}:
\begin{equation}
\begin{small}
\begin{aligned}
-\alpha D \partial_t q_{x} - \partial_t q_{y} &= -f_x^p - f_x^s, \\
\partial_t q_{x} - \alpha D \partial_t q_{y} &= -f_y^p - f_y^s.
\label{Thiele}
\end{aligned}
\end{small}
\end{equation}
Here, $ D \approx 1 $ stands for the (x,x) and (y,y) components of the dissipative force tensor  $ \mathbf{D}$:
$ \mathbf{D}_{ij} = D$ for $(i,j) = (x,x)$ and $(i,j)=(y,y)$, while $ \mathbf{D}_{ij} = 0$ otherwise~\cite{seidel}.
In turn,  $ f^p_{x,y} $ is the force acting on a particular skyrmion due to
the corresponding confining potential $ V $,  $ \vec{f}^p = -\vec{\nabla} V $.  $ f^p_{x,y} $ and $ V $ are determined from $ F_M $, see the supplementary information \cite{supp}. We calculate numerically the free energy and fitted it to the analytical
result obtained for the quadratic confinement function. The confining potential describes the effect of coupling between the
skyrmion and the corresponding nanodisk, and generally may be
different for different skyrmions. In particular,  we consider $ V = c r^2 $,
where $ r^2 = q_x^2 + q_y^2 $ and the pinning center is set as
$ (x,y) = (0,0) $.  Additionally, the neighbouring skyrmions
experience a repulsive interaction.
As the skyrmions are confined in the areas below the nanodisks,
the distance between them can vary only over a small range. Thus, one may
assume that the repulsive force in the confinement region is
constant.
Accordingly, the energy corresponding to the repulsion of two ($i$-th
and $j$-th) neighbouring skyrmions can then be written as
$ E_c = -c_u r_d $ (the detailed definition of coupling constant $ c_u >0$ and the distance between skyrmions $r_d$ see in the supplementary information \cite{supp}). The repulsion force acting
on the $i$-th skyrmion is determined as
$ f^{i,s}_{x,y} = -\partial E_c / \partial q^i_{x,y} $ (and similarly
for the $j$-th skyrmion). Note, the absolute value of the force acting
on the skyrmions is equal to $c_u$ and is measured in the units of m/s
(because the force in Thiele's equation is normalized to $M/\gamma $).
In what follows the
coefficients $c$ and $ c_u $ are phenomenological constants and are
tuned to achieve good agreement of the results based on Thiele's
equation and on the micromagnetic simulations.
To explore the role of the skyrmion-skyrmion interaction we first describe
the skyrmion precession in the absence of inter-skyrmion coupling,
($ c_u = 0 $), and adopt the ansatz $ q_x = q_{x0} \exp(i \omega t) $
and $ q_y = q_{y0} \exp(i \omega t) $. From the Thiele's equation we
derive the equation for the column vector
$ \vec{q} = (q_{x0}, q_{y0})^T $,
$ \hat{H} \vec{q} = \omega \vec{q} $. The explicit form of the matrix
$ \hat{H} $ reads:
\begin{equation}
\begin{small}
\displaystyle \hat{H} = \frac{1}{1+\alpha^2} \left( \begin{matrix}  -2 i \alpha c  & 2 i  c \\  -2 i  c & -2 i \alpha c \end{matrix} \right).
\label{ham}
\end{small}
\end{equation}
The eigenfrequencies of the matrix $ \hat{H} $  read: $ \omega_{\pm} = \pm\frac{2 c (1 \pm i \alpha)}{1 + \alpha^2}$.
The eigenvectors corresponding to the eigenvalues $ \omega_{+} $ and
$ \omega_{-} $ have the forms $ (-i, 1) $ and $ (i, 1) $,
respectively. The real parts of the eigenfrequencies correspond to
the skyrmion precession frequencies, and the imaginary parts describe
the attenuation. As one can see  in Fig. \ref{thiele}(a), the
coefficient of the confinement potential $ c $ is positive for a
negative field $ H_b $. Therefore, the steady clockwise precession is
described by the frequency $\omega_{+}$ and the corresponding vector
$ (-i, 1) $.
In turn, for positive field, $ H_b>0 $, the parameter $c$ is negative with a larger absolute value. The corresponding frequency $ \omega_{-} $ has larger real part as well and the vector
$ (i, 1) $ corresponds to counter-clockwise precession. This has been also confirmed in numerical simulations.
To achieve a good agreement with the results of micromagnetic simulations, we adjusted
the value of $ c $ for different values of $ H_b $, and calculated the
time dependence of the precession phase and total oscillations, as plotted
in Fig. \ref{thiele}. The analytical results are in good agreement
with those obtained from micromagnetic simulations, see
Fig. \ref{echo}.
 To analyze the influence of coupling between skyrmions, we explored the skyrmion echo as a function of the distance $d_s$ between skyrmions (in the experiment $d_s$ is equal to the distance between nanodisks).
	\begin{figure}
		\includegraphics[width=0.45\textwidth]{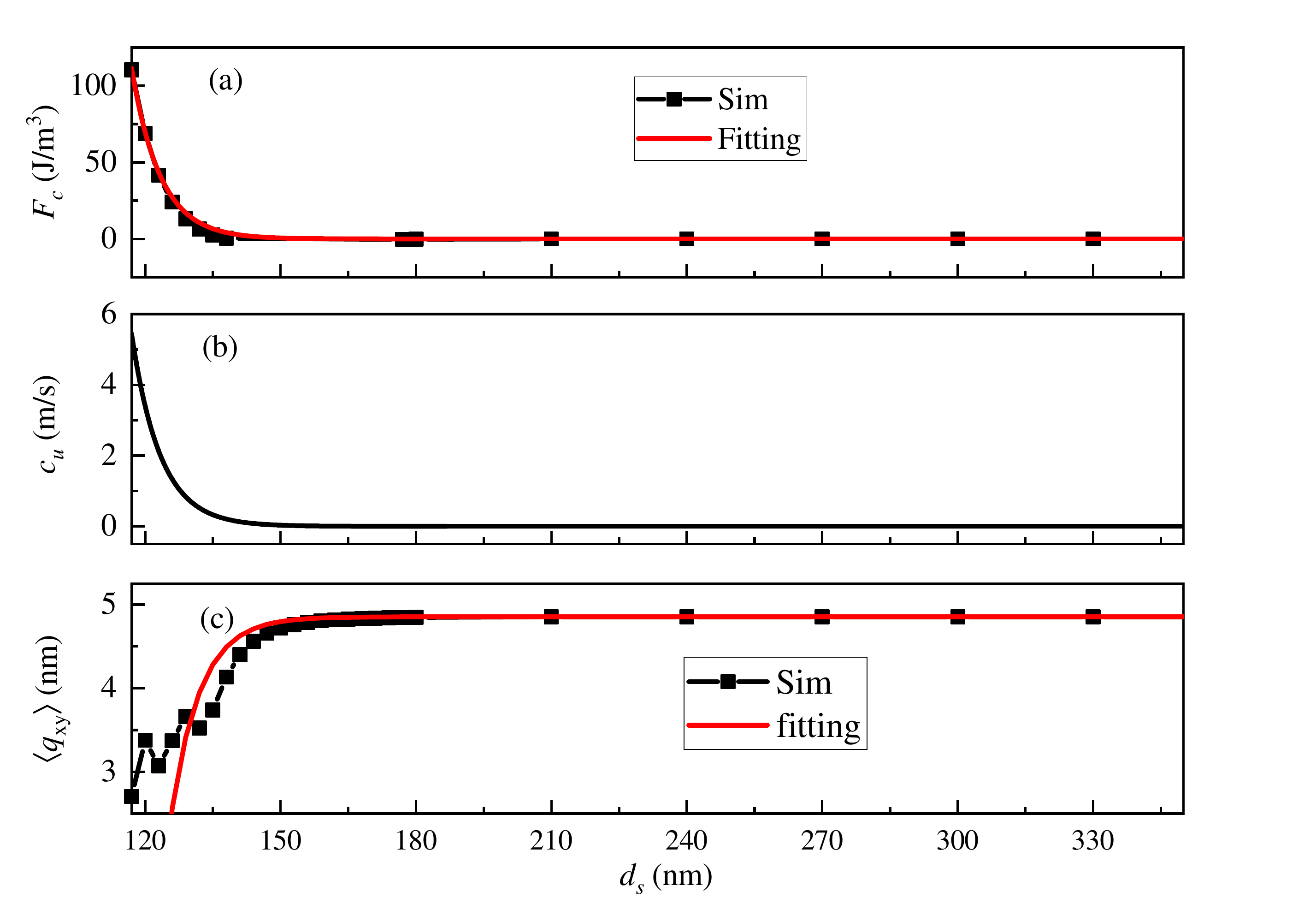}
		\caption{\label{distance}  (a) Dependence of the magnetic free energy density $ F_c $ on the distance $ d_s $ between neighboring skyrmions. The magnetic free energy is extracted from the simulation results. The energy $ F_c (d_s) $ fits to the exponential function $ F_c = A_0 \exp(-d_s/\xi) $ with the decaying length $ \xi = 6.4 $ nm. (b) The coupling strength $ c_u $ extracted from linear  expansion of  the energy profile in the vicinity of the skyrmion center. (c) The amplitude $ \langle q_{xy} \rangle $  of the skyrmion echo as a function of the distance $ d_s $ between neighboring skyrmions. Amplitudes are fitted to the exponential function $ \langle q_{xy} \rangle = I_1 - I_0 \exp(-d_s / \xi) $.} The bias field is $ H_z = 100 $ mT.
	\end{figure}
The coupling strength between skyrmions increases with decreasing $ d_s $ \cite{leonov2015multiply}. As shown in Fig. \ref{distance}, the coupling energy and coupling strength exponentially decay for large $d_s$, while the echo signal increases and saturates. When decreasing $ d_s $, the coupling strength increases and the echo signal $ \langle q_{xy} \rangle $ becomes reduced. The observed effect has a clear physical explanation:  For the short $ d_s $, the observed non-monotonic behavior is related to collective oscillations caused by the strong coupling between skyrmions. For $d_s>150$nm, the interaction between skyrmions is weak, and the echo signal is insensitive to the distance between skyrmions, see Fig. \ref{distance}. One can extract information on the skyrmion-skyrmion interaction strength by performing echo experiments for $d_s<150$nm.\vspace{0.1cm}\\
\textit{Summary and conclusions}:
 The inhomogeneus field leads to a dephasing of an
initial signal. Switching the magnetization of the nanodisks (e.g due to the application of a spin-polarized torque)
turns the dephasing process into a rephasing one, and after a certain time, the signal of the skyrmion echo is recovered. The proposed skyrmion echo is experimentally
feasible and can be detected  by exploiting the spin pumping and inverse spin Hall
effects. The skyrmion echo will also be important for the coupled systems of skyrmions and superconducting vortexes
  \cite{PhysRevLett.122.097001}, i.e. for a hypothetical superconducting vortex echo.\vspace{0.1cm}\\
\textit{Acknowledgements}: The work is supported by Shota Rustaveli
National Science Foundation of Georgia (SRNSFG) (Grant
No. FR-19-4049), the National Natural Science Foundation of China
(Grants No. 12174452, No. 11704415 and No. 12074437), the Natural Science Foundation
of Hunan Province of China (Grants No. 2022JJ20050 and No. 2021JJ30784), and by the National
Science Center in Poland by the Norwegian Financial Mechanism
2014-2021 under the Polish-Norwegian Research Project NCN GRIEG
(2Dtronics) no. 2019/34/H/ST3/00515 (AD,JB), the FWF International Project
I 5384, and as a research Project No. DEC-2017/27/B/ST3/ 02881 (VKD).

\section{Supplementary Information}
\beginsupplement
\subsection{Numerical simulation details}
The LLG equation is numerically solved employing a fifth-order Runge-Kutta scheme with a fixed time step of 0.5 ps. We adopted the finite difference approximation, and discretized the ferromagnetic layer in the unit simulation cells $ s \times s \times s$ with $ s = 3 $ nm. To quantitatively characterize the skyrmion structure, we used the skyrmion topological charge density $ c = (1/4\pi) \vec{m} \cdot (\partial_x \vec{m} \times \partial_y \vec{m}) $, and the total topological charge $ C = \int d^2 \vec{r} \, c $.  The position of $i$-th skyrmion centre,  $ \vec{q}_i = (q_x^i, q_y^i) $, is weighed by the topological charge: $ \vec{q}_i = \int_i d^2 \vec{r}\, \vec{m} \cdot (\partial_x \vec{m} \times \partial_y \vec{m})\, \vec{r} / \int d^2 \vec{r} \, \vec{m} \cdot (\partial_x \vec{m} \times \partial_y \vec{m}) $ \cite{PhysRevLett.111.067203}. For each skyrmion the integration range is limited by the area near the corresponding pinning center. When integrating over the whole magnetic layer one finds $ \sum_{i=1}^{n} \vec{q}^i $.

\begin{figure}
	\includegraphics[width=0.48\textwidth]{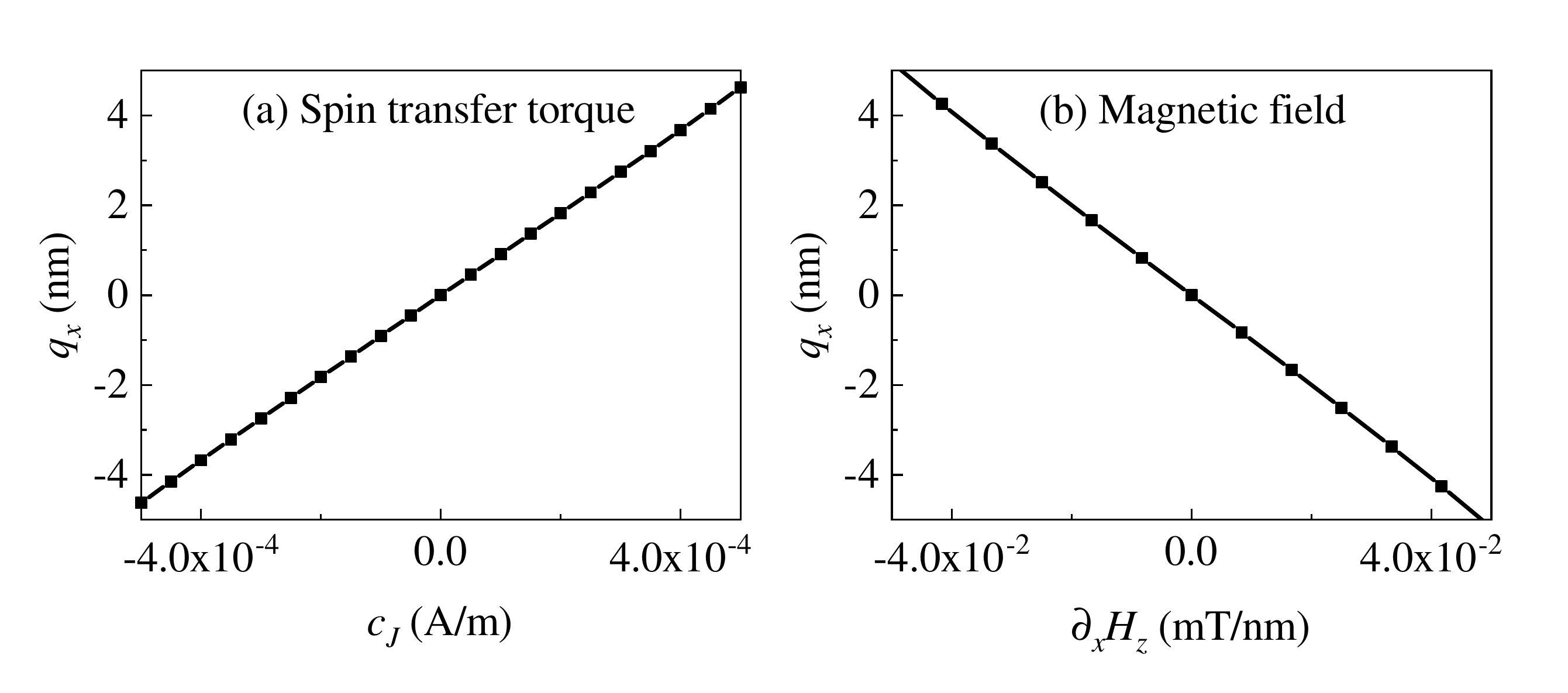}
	\caption{\label{initial} The initial displacement of a skyrmion,  $ q_x $, along the $ \textbf{x} $ axis,  driven by spin transfer torque (a) or spatially varying external magnetic field (b).}
\end{figure}

To shift slightly the initial positions of skyrmions  one can apply a spin-transfer torque \cite{Tomasello2014} or a magnetic field \cite{PhysRevB.92.064412}. For example, the applied spin transfer torque $ \vec{\tau}_{\rm STT} = \gamma c_J \vec{m} \times \vec{y} \times \vec{m}, $ with the electron polarization along the $ y $ axis, steers the skyrmion center in the $ x $ direction. As shown in Fig. \ref{initial}(a), positive (negative) $ c_J $ shifts the skyrmion along $ +(-)x $ direction, and the induced displacement $ q_x $ depends linearly on $ c_J $ (or electric current density). The same can be achieved by spatially inhomogeneous magnetic field $ H_z $. The gradient of the field, $ \partial_x H_z $, linearly shifts the skyrmion along the axis $ \pm x $ and the sign depends on the sign of the gradient $ \partial_x H_z $, as it is shown in Fig. \ref{initial}(b).
\begin{figure}
	\includegraphics[width=0.48\textwidth]{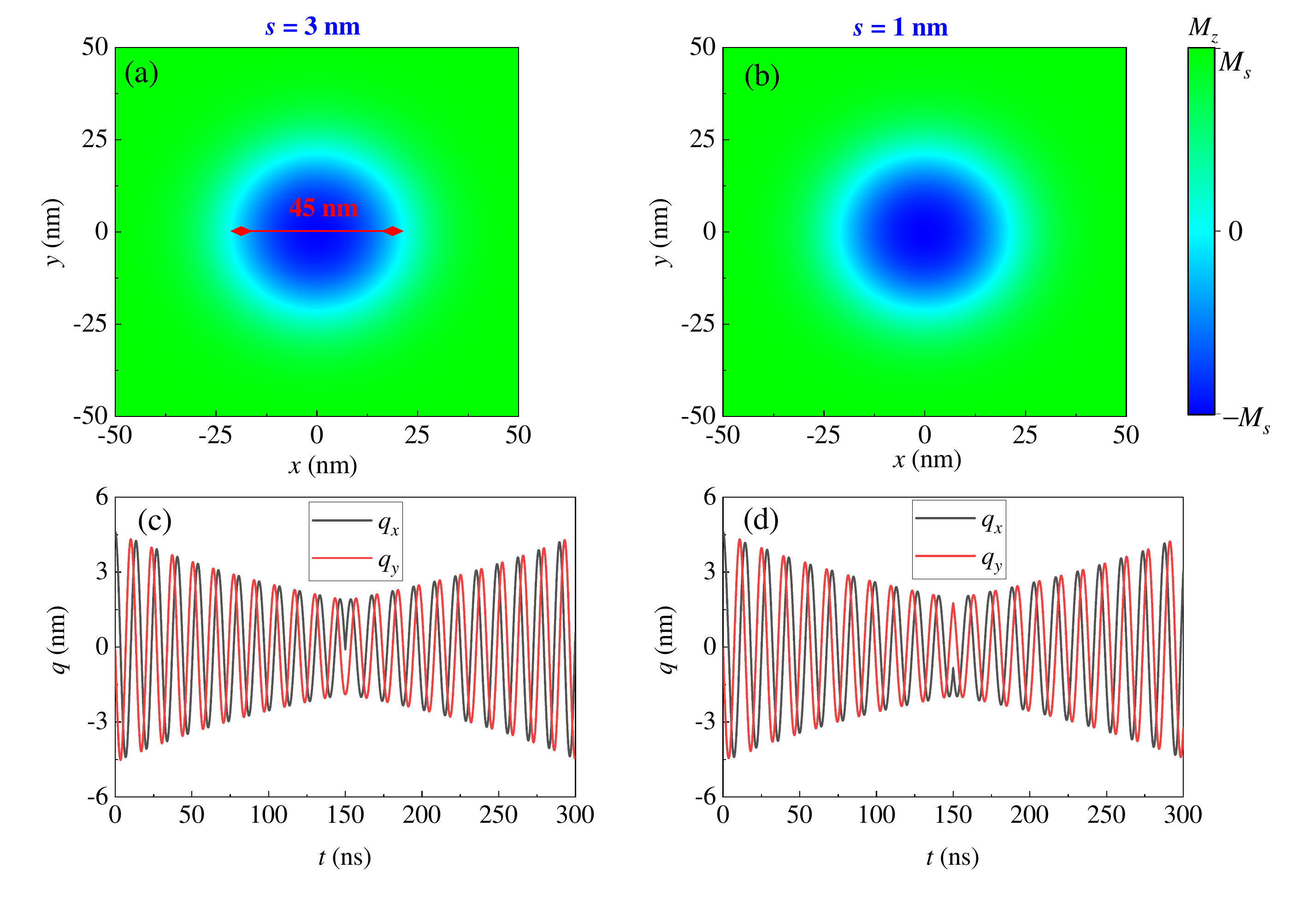}
	\caption{\label{cellsize} (a,b)The $ M_z $ component of the skyrmion magnetic texture for two different values of cell size: $ s = 3 $ nm (a) and $ s = 1 $ nm (b). (c,d) Precession of the  single skyrmion around the center $ (0, 0) $ of the nanodisk for $ s = 3 $ nm (c) and $ s = 1 $ nm (d). At $ t = 150 $ ns, the direction of the coupling field $ H_b $ is changed from $ -15 $ mT to $ 15 $ mT. The bias magnetic field $ H_z = 100 $ mT and the size of the stable skyrmion is about 45 nm. }
\end{figure}
\begin{figure}
	\includegraphics[width=0.48\textwidth]{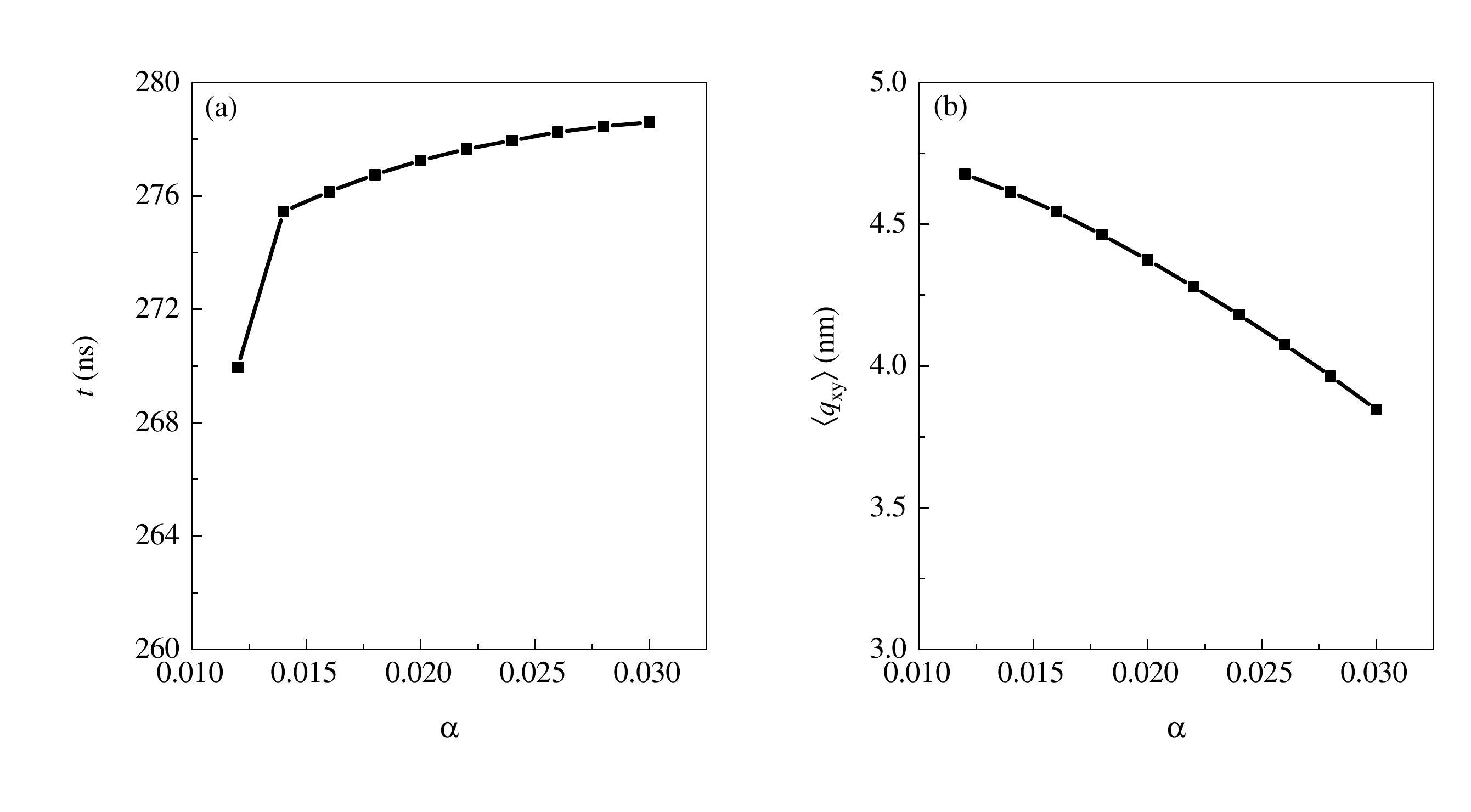}
	\caption{\label{damp} (a) Dependence of the rephasing time on the damping constant $ \alpha $. For a stronger damping, the  rephasing process becomes slower. (b) The amplitude of skyrmion echo $ \langle q_{xy} \rangle = \frac{1}{n} [(\sum_{i=1}^{n}q_x^i)^2 + (\sum_{i=1}^{n}q_y^i)^2 ]^{1/2} $ as a function of the damping constant $ \alpha $. The bias field is assumed $ H_z = 100 $ mT.  The direction of the coupling field in all nanodisks is reversed at $ t = 150 $ ns, and the distance between neighboring skyrmions is $ d_s = 270 $ nm.}
\end{figure}

To exclude numerical artifacts of the coarse-graining procedure, we performed calculations for the smaller size of the cell, $ s = 1 $ nm. As it is shown in Fig. \ref{cellsize}, the obtained result is almost identical with that obtained for $ s = 3 $ nm. The difference in precession frequencies is about $ 5 \% $. The radius $ 45 $ nm  of the skyrmion is larger than the radius of the nanodisk $ 12 $ nm.  Confinement of the skyrmion by nanodisk is quite efficient.

We have also analyzed the impact of the Gilbert damping constant $ \alpha $ and found that the magnitude of damping parameter $ \alpha $ has a significant influence on the skyrmion relaxation and on the rephasing of the echo, see Fig. \ref{damp}. The enhanced damping slows down the rephasing process and also decreases the amplitude of the skyrmion echo.

 In the main text, the averaged magnetic energy density is calculated from the formula $ F_M = -\frac{\mu_0 M_s}{S} \int \vec{m} \cdot \vec{H}_{\rm eff} d^2 r  $ applied to the region inside of the skyrmion (cross-section $ S $). The spatial profile of the magnetization vector $ \vec{m} $ and the effective field $ \vec{H}_{\rm eff} $ are obtained from simulation results. To obtain the position-dependent $ F_M $ curve (Fig. 3(a) in the main text), we fixed the initial stable skyrmion texture and gradually moved the pinning center along the $ x $ axis.
	Then, from the spatial gradient of $ F_M $, one obtains the real phenomenological confining force  acting on the skyrmion. However, due to specific normalization of the Thiele equation, to implement the confining force into this equation one needs to normalize accordingly the confining potential. Therefore,
	we introduce the relevant potential $V$ as $ V = c_p F_M = (c_p A) r^2 = c r^2 $, and in
	Fig. 1(c) in the main text) we used $ c_p = 0.31 \times 10^{-9} {\rm m^5/Js} $. The confining force $ f^p_{x,y} $ is the give by the gradient of $V$, $ f^p_{x,y} =- \partial V/\partial x,y $.

\subsection{Different bias field and distance}

\begin{figure}
	\includegraphics[width=0.48\textwidth]{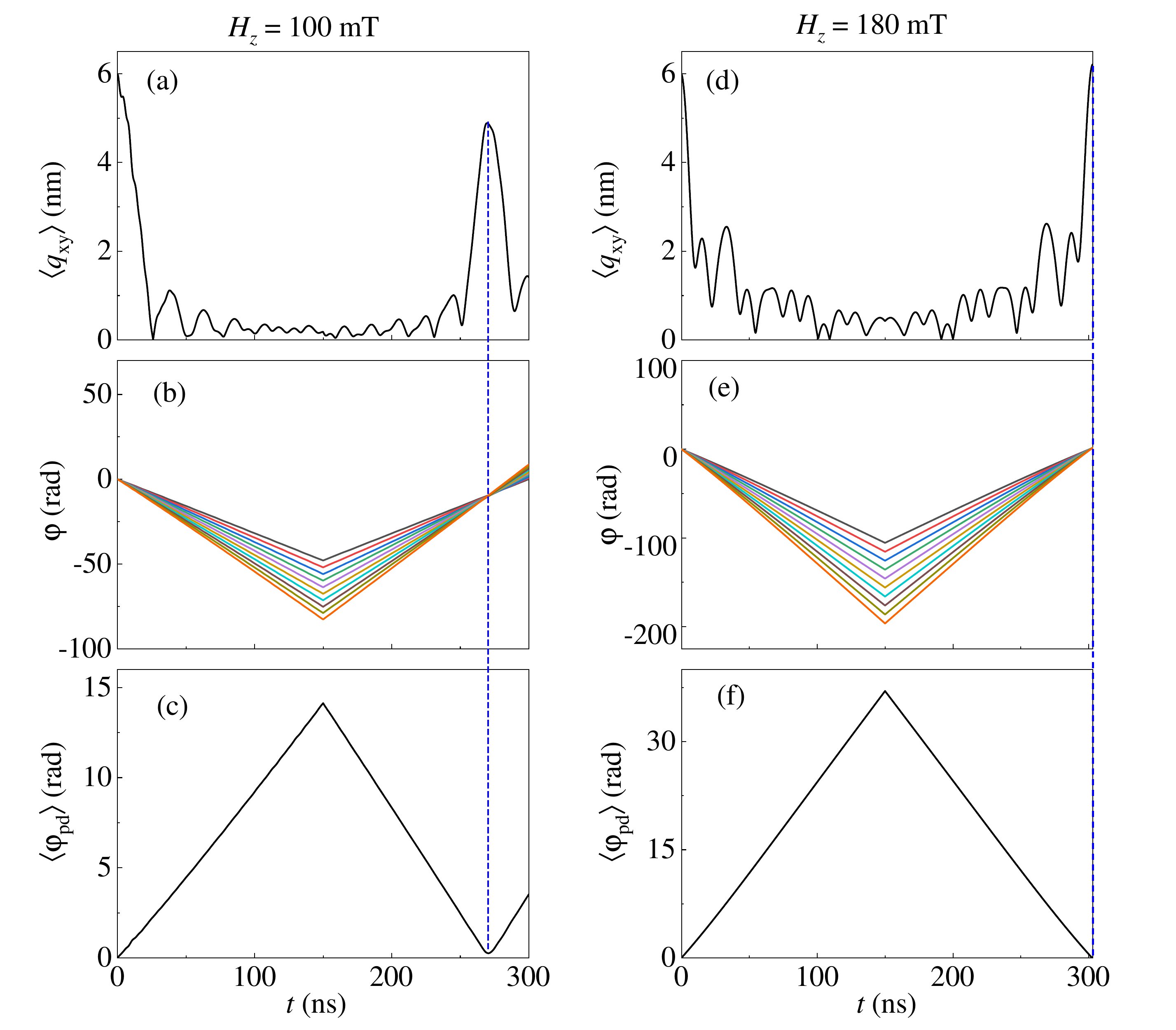}
	\caption{\label{difffield} (a,d)  Time dependence of the amplitude of skyrmion echo averaged over all  (n=10) skyrmions, $ \langle q_{xy} \rangle = \frac{1}{n} [(\sum_{i=1}^{n}q_x^i)^2 + (\sum_{i=1}^{n}q_y^i)^2 ]^{1/2} $, plotted for two different bias fields $ H_z = 100 $ mT and $ H_z = 180 $ mT.
		(b,e) Evolution of the precession phases of all $n=10$ skyrmions. (c,f) The averaged phase differences between the $n=10$ skyrmions, evaluated as $ \langle \varphi_{pd} \rangle = \sum_{i=1}^{n}\sum_{i=j}^{n}(\varphi_j - \varphi_i) / n^2 $. The direction of the coupling field in all nanodisks is reversed at  $ t = 150 $ ns, and the distance between neighboring skyrmions is $ d_s = 270 $ nm.}
\end{figure}
\begin{figure}
	\includegraphics[width=0.48\textwidth]{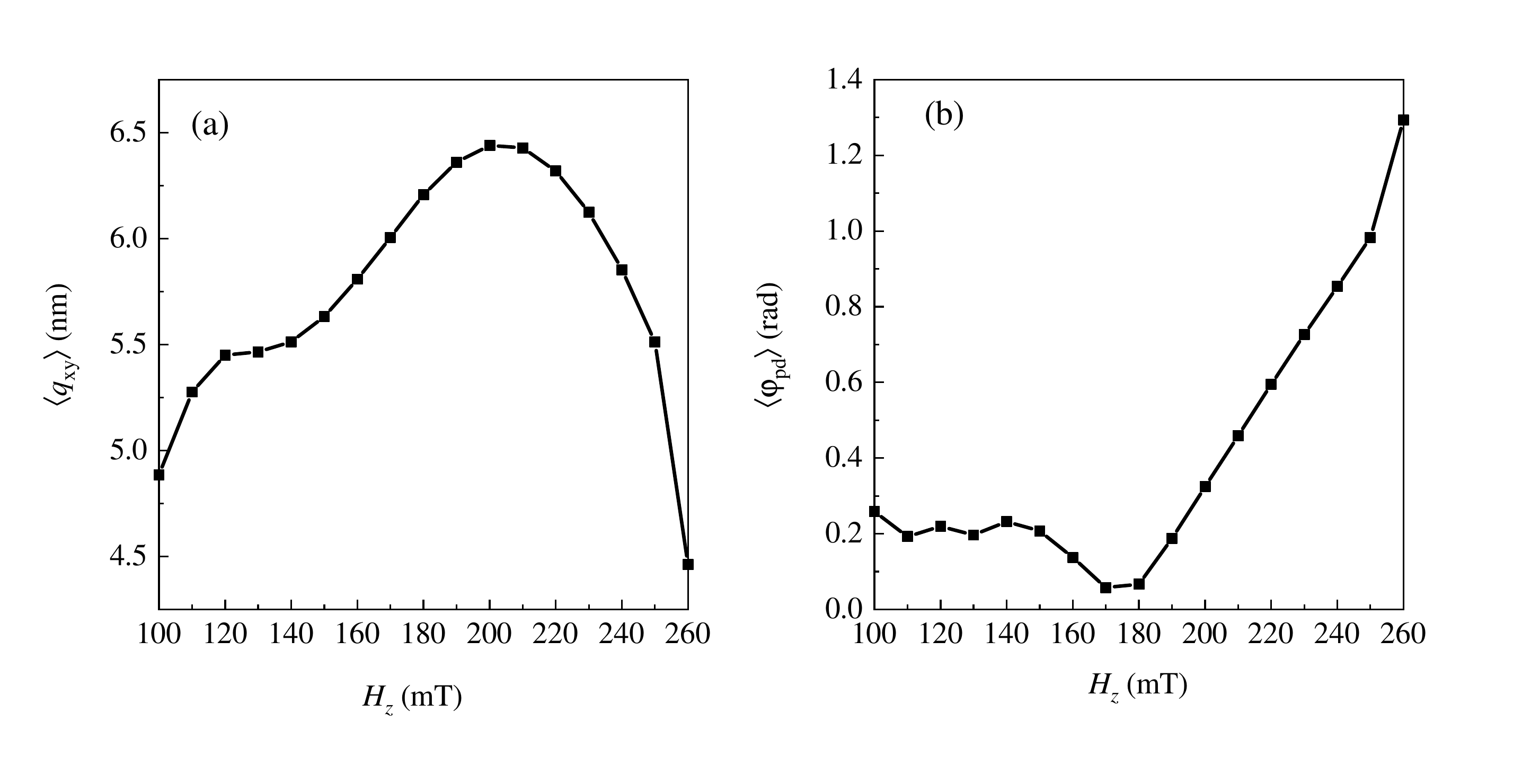}
	\caption{\label{varyfield} (a) The amplitude $ \langle q_{xy} \rangle $ and (b) minimum phase difference  $ \langle \varphi_{pd} \rangle $ of the skyrmion echo as a function of the bias field $ H_z $. The distance between neighboring skyrmions is $ d_s = 270 $ nm. }
\end{figure}

The dephasing and rephasing times are slightly different. This fact can have an impact on the signal of the skyrmion echo. Therefore it is necessary to evaluate the quality of rephasing of the skyrmion echo. Let us define average difference between the skyrmion phases through the equation  $ \langle \varphi_{pd} \rangle = \sum_{i=1}^{n}\sum_{i=j}^{n}(\varphi_j - \varphi_i) / n^2 $.  In the case of exact rephasing one finds $ \langle \varphi_{pd} \rangle = 0 $.  In Fig. \ref{difffield} we see that signal of the skyrmion echo appears when $ \langle \varphi_{pd} \rangle $ is minimal. For the bias field $ H_z = 100 $ mT, the minimum of  phase differences is $  \langle \varphi_{pd} \rangle = 0.25$ at $ t = 270 $ ns. With increasing $ H_z $, the asymmetry between dephasing and rephasing times gradually decreases.
At $ H_z = 180 $ mT, the minimum of $ \langle \varphi_{pd} \rangle \approx 0 $ for $ t = 300 $ ns indicates on a perfect rephasing, and the amplitude of the corresponding echo signal becomes larger.  With a further increase of $ H_z $, the rephasing time becomes larger and the minimum of $ \langle \varphi_{pd} \rangle $ increases again, as is shown in Fig. \ref{varyfield}.  The amplitude of the skyrmion echo, $ \langle q_{xy} \rangle $, increases until $ H_z = 210 $ mT, while $ \langle \varphi_{pd} \rangle $ is not zero. The difference between behavior of $ \langle q_{xy} \rangle $ and $ \langle \varphi_{pd} \rangle $ is related to the bigger rephasing time. The skyrmion trajectory becomes larger in the rephasing stage. For example, at $ H_z = 210 $ mT, the echo (minimum of $ \langle \varphi_{pd} \rangle $) is achieved at $ t = 330 $ ns, and  larger rephasing time leads to the greater amplitude $ \langle q_{xy} \rangle $.

Furthermore, as discussed in the main text, the skyrmion-skyrmion coupling strength $ c_u $ also affects the skyrmion echo strength. We suggest varying the distance $ d_s $ between neighboring skyrmions (and nanodisks) to reproduce the experimentally feasible small dispersion of the coupling strength. According to Ref. \cite{Leonov2015, PhysRevB.87.214419, PhysRevLett.119.207201, PhysRevLett.125.227201}, the coupling strength between skyrmions increases with decreasing distance $ d_s $.  In the analysis, the coupling strength $ c_u $ is defined from the linear ansatz. First, from the numerical simulation  we extracted the magnetic free energy density $ F_c $ as a function of the distance $ d_s $ (also the distance between two neighboring skyrmions), which fits to the exponential function $ F_c = A_0 \exp(-d_s/\xi) $. For application in the  Thiele equation,  the repulsion energy is given by the formula, $ E_c = c_p A_0 \exp(-r_d/\xi) $, where $ r_d = \sqrt{(d_s + q_x^i - q_x^j)^2 + (q_y^i - q_y^j)^2} $ is the distance between two skyrmions. From $ f_{x,y}^{i,s} = -\partial E_c / \partial q_{x,y}^i $, we find the repulsion forces $ f_{x}^{i,s} = E_c (d_s + q_x^i - q_x^j) /\xi r_d $ and $ f_{y}^{i,s} = E_c (q_y^i - q_y^j) /\xi r_d $. As the skyrmion oscillates around the pinning center with a small amplitude, in this small range  we adopted the linear ansatz $ E_c = -c_u r_d $. The forces can be then rewritten as   $ f_{x}^{i,s} = c_u (d_s + q_x^i - q_x^j) /r_d $ and $ f_{y}^{i,s} = c_u (q_y^i - q_y^j) /r_d $. The coupling strength is determined through comparing two expressions  $ c_u = E_c/\xi = c_p A_0 \exp(-r_d/\xi)/\xi  $.

\begin{figure}
	\includegraphics[width=0.45\textwidth]{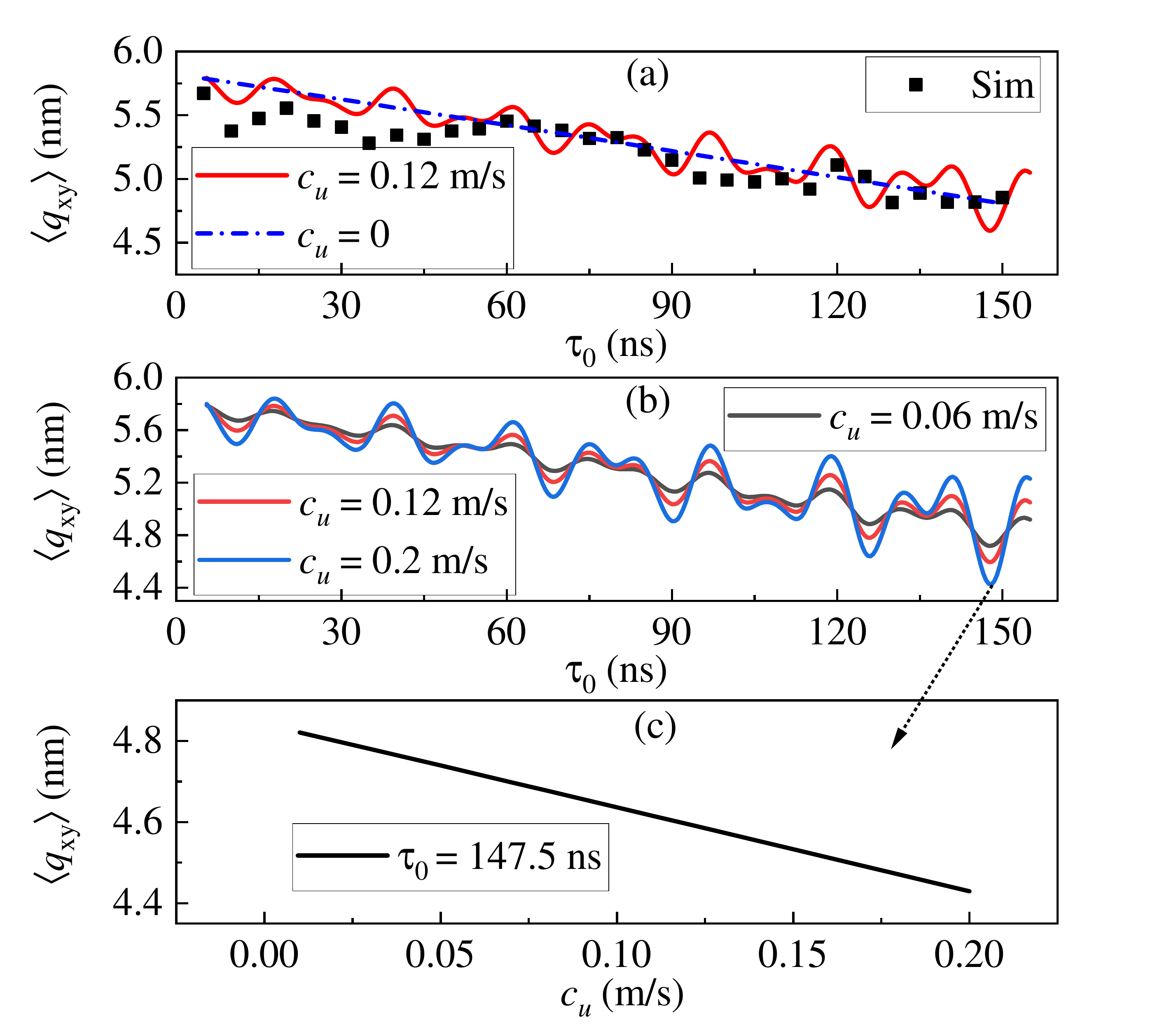}
	\caption{\label{coup}\color{black} (a) Amplitude of the skyrmion echo {\color{black} $ \langle q_{xy} \rangle = \frac{1}{n} [(\sum_{i=1}^{n}q_x^i)^2 + (\sum_{i=1}^{n}q_y^i)^2 ]^{1/2} $} as a
		function of pulse duration $ \tau_0 $. Black squares correspond to
		the micromagnetic simulations, whereas the blue dashed line (the
		coupling strength $ c_u = 0 $) and the red solid circles (the
		coupling strength $ c_u = 0.12 $ m/s) are obtained from Thiele's
		equation. (b) Numerical solution of Thiele's
		equation, for different values of $ c_u $.  (c)
		Dependence of the amplitude of skyrmion echo on the strength of
		inter-skyrmion coupling $ c_u $ for $\tau_0 = 147.5 $
		ns.
	}
\end{figure}

An interesting question is the change of the echo strength with the pulse duration
$ \tau_0 $.  From the simulation results shown in Fig. \ref{coup}(a),
one can see that the amplitude of the echo decreases with the duration
of the pulse $ \tau_0 $ and slightly fluctuates. We analyze this
dependence by the numerical solution of Thiele's equation. When all skyrmions are independent ($ c_u = 0 $), the
echo strength decreases monotonically with $\tau_0 $ (Fig. \ref{coup}(a)). A strong enough coupling ($ c_u = 0.12 $ m/s)
leads to fluctuations similar to those observed in the micromagnetic simulations. With the increase of the skyrmion-skyrmion interaction
$ c_u $, the fluctuation amplitude increases, but the shape of the
curve remains unchanged.

We note that opposite to the conventional spin-echo, the trajectories
of the skyrmions are helixes and therefore for individual
skyrmions
$ \langle q_{xy} \rangle = [(q_x^i)^2 + (q_y^i)^2 ]^{1/2} $ are
not circles of constant radius \cite{PhysRevLett.125.227201}. This difference should be taken into account in order to understand behavior of the skyrmion echo amplitude with the coupling strength and pulse duration.

\subsection{Influences of demagnetization field}
For a more realistic discussion, we analyze the influences of dipole-dipole interaction on the skyrmion echo. In the simulation, we also include the uniaxial anisotropy field $ \frac{2 K_u}{ \mu_0 M_s} m_z \vec{z} $ along the $ z $ axis with the constant $ K_u = 0.905 \times 10^6 {\rm Jm^{-3}} $, as well as the demagnetization field,
\begin{equation}
\begin{small}
\begin{aligned}
\vec{H}_{\rm demag}(\vec{r}) = -\frac{M_s}{4 \pi} \int_V \nabla \nabla' \frac{\vec{m}(\vec{r}')}{|\vec{r} - \vec{r}'|} d\vec{r}'.
\label{demag}
\end{aligned}
\end{small}
\end{equation}

\begin{figure}
	\includegraphics[width=0.48\textwidth]{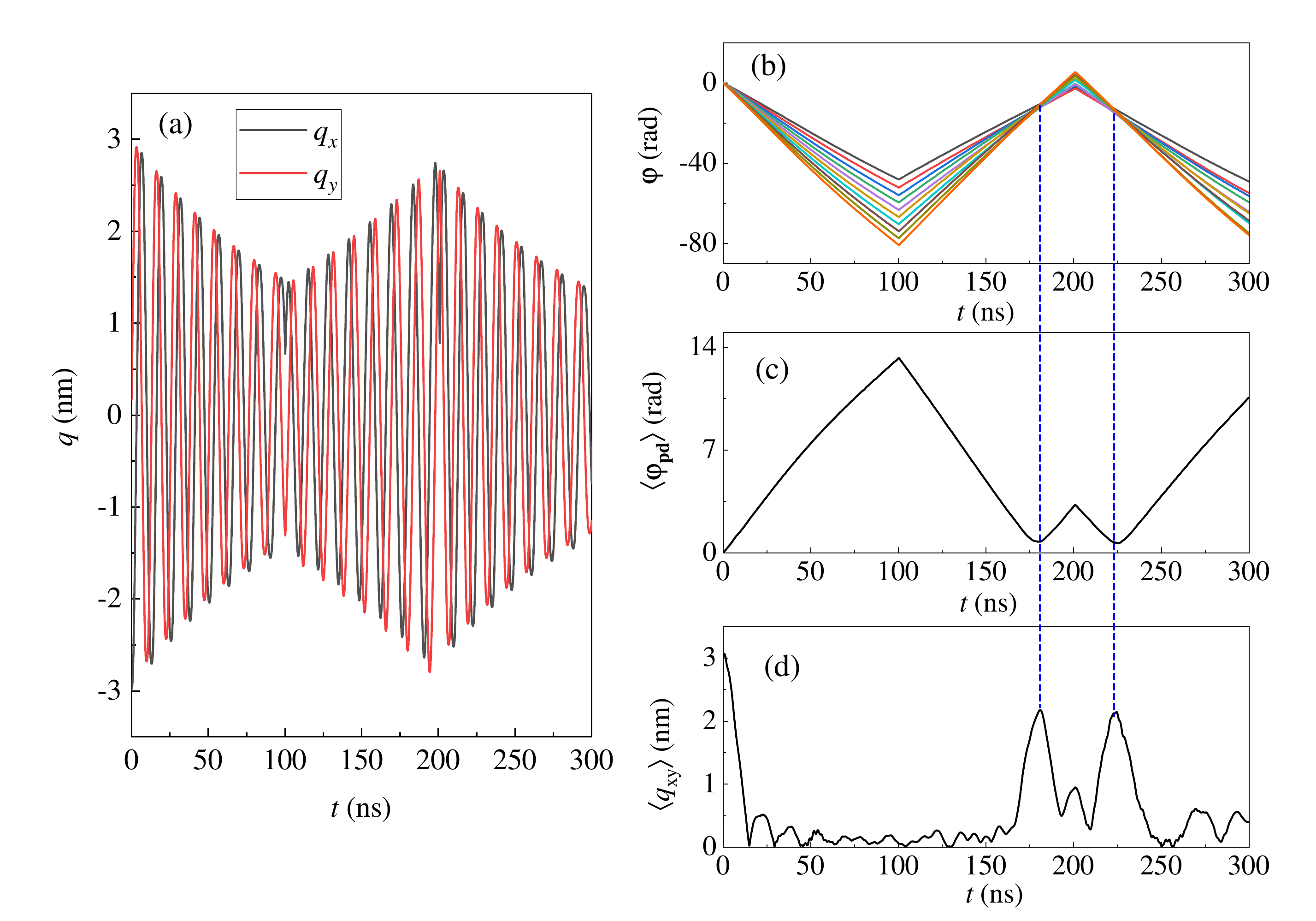}
	\caption{\label{demagf} The effect of the demagnetization field. (a) Dynamics of single skyrmion precession around the center $ (0, 0) $ of the nanodisk.  (b-d) Time evolution of the precession phases, phase difference $ \langle \varphi_{pd} \rangle $, and averaged $ \langle q_{xy} \rangle $ over ten skyrmions calculated from micromagnetic simulations. The direction of coupling field $ H_b $ in the nanodisk is reversed periodically after {\color{black} 100 ns}, and the distance between neighboring skyrmions (and thus nanodisks) is $ d_s = 540 $ nm.}
\end{figure}

The film is uniformly magnetized along the $ z $ axis and the effective field of uniaxial anisotropy  $ 1.2 \times 10^6 \rm{A/m} $ is partially compensated by the static demagnetization field $ -N_z M_s $, with the film demagnetization factor $ N_z =1 $. We found a slight effect of the dipole-dipole interaction on the skyrmion dynamics. The main conclusions concerning the skyrmion echo still hold. The results for the skyrmion echo in the system with the demagnetization field are shown in Fig. \ref{demagf}. We again studied the same system of $ n = 10 $ separated skyrmions pinned under ten nanodisks with slightly different coupling fields, and the distance between neighboring nanodisks was $ d_s = 540 $ nm.   One can clearly see the skyrmions dephasing and rephasing, minimum in the phase difference $ \langle \varphi_{pd} \rangle $, and the skyrmion echo.

\bibliography{2nems}

\end{document}